\pdfoutput=1
\documentclass{elsart}

 \usepackage{graphicx}
\def\bra#1{\langle #1|}
\def\ket#1{|#1 \rangle}

\usepackage{amssymb, amsmath}

\begin{document}

\begin{frontmatter}


\title{Impact of Electron Density on the Fixed-Node Errors in Quantum Monte Carlo}

\author{K. M. Rasch\corauthref{cor1}}
\author{L. Mitas}
\ead{kmrasch@ncsu.edu}
\address{CHIPS, Department of Physics, North Carolina State University, Raleigh, North Carolina 27695, USA}
\ead{lmitas@ncsu.edu}
\corauth[cor1]{}

\begin{abstract}  We analyze the effect of increasing charge density on the Fixed Node Errors in Diffusion Monte Carlo by comparing FN-DMC calculations  of the total ground state energy on a 4 electron system done with a Hartree-Fock based trial wave function to calculations by the same method on the same system using a Configuration Interaction based trial wave function.  We do this for several different values of nuclear charge, Z.  The Fixed Node Error of a Hartree-Fock trial wave function for a 4 electron system increases linearly with increasing nuclear charge.\end{abstract}

%\begin{keyword}

%\PACS 
%\end{keyword}
\end{frontmatter}

% main text
\section{Introduction}
\subsection{place of QMC amongst other electronic structure models}
Quantum Monte Carlo (QMC) represents a promising
methodology for solving electronic structure and quantum many-body problems in general.
The common thread of QMC methods is the use of stochastic sampling both to evaluate 
expectation values for many-body wave functions as well as to solve the stationary Schr\"odinger equation 
\cite{mi}.
In addition to being a useful tool for treating interactions in a many-body framework, 
QMC methods can handle many different systems ranging from atoms to molecules to extended crystals.
The obtained results for energy differences are typically within a few percent of experimental values.  Even 
on absolute scale QMC correlation energies are remarkably accurate, reaching 90-95\% of the exact values. 
The favorable scaling in the number of particles makes QMC very promising both in applications 
and also in further theoretical developments.

The key approximation which hampers further increase in accuracy is the so-called fixed-node
approximation which is used to overcome the notorious fermion sign problem.  
The fermionic antisymmetry implies that wave functions will have both positive and negative negative 
values; however, the statistical approaches such as QMC typically require non-negative distributions
otherwise they become inefficient.
One means of overcoming this fundamental difficulty is to use the node (wave function zero locus)
between the positive and the negative regions as a boundary condition in solving the Schr\"odinger equation.
This avoids the sign problems since one finds the absolute value of the wave function
with prescribed boundaries.  In the case that one would know the exact nodal hypersurface, the method would provide 
the exact solution. Unfortunately one seldom possesses such knowledge.  Therefore we have to use the best 
available nodes which come from optimized trial (or variational) wave functions.  We fix
the nodal surface of the solution to be identical to the nodes of the trial function and solve the
Schr\"odinger equation with approximate nodes. The fixed-node approximation provides an upper bound 
to the true energy, and in practice it provides very remarkable accuracy as mentioned above.

The goal of this work is to study the impact of increasing electronic density on the fixed-node bias. 
This requires a system which shows significant error in the mean-field (Hartree-Fock)
 nodes but for which we know an excellent approximation to the exact node \cite{br,um,um2} 
which is not too difficult to construct across the range of densities. 
In particular, we focus on studying a ``Be-like'' series of four electron systems in a 
Coulombic potential with varying nuclear charge $Z=3$ to $Z=28$. We contrast the fixed-node 
bias of the Hartree-Fock vs. two-configuration trial wave functions and analyze the corresponding 
trends as $Z$ increases.

\subsection{Basics of DMC and the fixed-node approximation.}
 Let us briefly mention the key facts about the fixed-node diffusion Monte Carlo method (FN-DMC).
 Consider an operator
$exp(-\tau H)$ where $H$ is the Hamiltonian and $\tau$ is a real
parameter (imaginary time). It is straightforward to show 
that applying this operator to an initial trial wave function $\Psi_T$
\begin{equation*}
\lim_{\tau\to \infty} \exp(-\tau H)\Psi_T = \Psi_0
\end{equation*}  
projects out the ground state $\Psi_0$ within the given symmetry class. 
This projection can be formulated as a solution of the imaginary-time Schr\"odinger equation 
\begin{equation*}
f(\mathbf{R},\tau + \tau') = \int G(\mathbf{R}\leftarrow\mathbf{R}', \tau') f(\mathbf{R},\tau) \label{diffusion}
\end{equation*}
where 
\begin{equation*}
G(\mathbf{R}\leftarrow\mathbf{R}', \tau')=\Psi_T(\mathbf{R}) \Psi_T^{-1}(\mathbf{R}')
\langle \mathbf{R}|e ^{-\tau H}|\mathbf{R}'\rangle 
\end{equation*}
is the Green function and $\mathbf{R}$ is the vector
of electron coordinates. We have introduced the importance sampling by $\Psi$ so that the desired
ground state $\Phi$ is obtained from the solution 
 $f(\mathbf{R},\tau)=
\Phi(\mathbf{R},\tau)\Psi_T(\mathbf{R})$ for large $\tau$.
Note that the fixed-node condition implies that $\Phi$ vanishes at the nodes of $\Psi_T$ and also that 
$f(\mathbf{R},\tau)\geq 0$ for any $\mathbf{R},\tau$.
Further details about the FN-DMC method can be found in Ref.~\cite{mi}. 
 
\subsection{Origin of nodal errors: topology of two versus four nodal domains in 4$e^-$ system.}
Let us consider the lowest state of four electrons in the Coulomb potential with charge Z (we assume 
atomic units throughout).
To aid in understanding the nature of fixed-node errors for this system, we recall the 
properties of the nodal domain topology 
of the two different trial wave functions.  The  
 Hartree-Fock (HF) trial wave function for Be is given by
 a Slater determinant which is block diagonal in spin so that it can be broken into a product of the spin channels
\begin{align*}
\Psi_{HF}(\mathbf{R})=&\det[\phi_{1s}(r_1),\phi_{2s}(r_2)] \times \det[\phi_{1s}(r_3),\phi_{2s}(r_4)]
\end{align*}
where $r_i=|\mathbf{r}_{i}|$.
  The nodes are defined implicitly for any trial wave function as
\begin{equation*}
\Psi_{T}(\mathbf{R}) =0
\end{equation*}
Let us analyze the determinant for 
the non-interacting orbitals first.  We can write analytically the hydrogenic functions as
\begin{equation*}
\phi_{1s}=e^{-Zr}
\end{equation*} 
and 
\begin{equation*}
\phi_{2s}=[1-(Zr/2)]e^{-Zr/2}
\end{equation*}
so that we get 
\begin{equation*}
h(r_1)h(r_2)[g(r_1) -g(r_2)]=0
\end{equation*}
where $h(r)$ is a function which does not vanish for any finite $r$ while
\begin{equation*}
 g(r)=\exp(-Zr/2) -1+(Zr/2)
\end{equation*}
It is then easy to show that the function $g(r)$ is monotonous since $g'(r)>0$
for any $r>0$. 
Using numerically accurate  HF orbitals for Be,
 it is possible to carry out similar arrangements
 and to demonstrate qualitatively the same behavior.
The monotonicity is important since it immediately implies that
equation $[g(r_1) -g(r_2)]=0$ is equivalent to $r_1 - r_2=0$ and also that 
this is the only solution. 
The node of the HF/non-interacting trial function is then given by
 \begin{equation*}
 (r_1 - r_2) (r_3 - r_4)=0,
\end{equation*}
which clearly shows that there are $2\times2=4$ nodal pockets~\cite{br}.  
However, it has been found some time ago that the correct number of nodal domains is two,
see, for example, Ref. ~\cite{br}. (For our purposes 
it is very instructive that the corresponding fixed-node bias in the correlation energy is significant,
about 10\% of the correlation energy
\cite{um}.)
The accurate nodal surface for the $^1S$ ground state is actually remarkably well described
by  two configurations wave function where HF is 
augmented by adding $2s^2\to2p^2$ double excitation which corresponds to 
a near-degeneracy effect. This configuration state function is a sum 
  of three Slater determinant products $\det[\phi_{1s},\phi_{2p_i}]\det[\phi_{1s},\phi_{2p_i}]$ 
for $i=x,y,z$ as required by
 the  $^1S$ symmetry. The 2-configuration wave function is thus given 
as 
 \begin{equation*}
 \Psi_{2conf}=d_0\Psi_{HF} +d_1\Psi_{2s^2\to2p^2} 
\end{equation*}
where the $d_0$ and $d_1$ are the variational coefficients.

\section{Trial wave function}
 We write the  many-body wave-function 
 as a product of an antisymmetric determinantal part, $\Psi_A$, times a Jastrow correlation factor
\begin{equation*}
\Psi_T(\mathbf{R}) = \Psi_A(\mathbf{R}) \cdot {\rm e}^{ J(\mathbf{R}) }
\end{equation*}
 Our Jastrow factor includes 1-, 2-, and 
 3-body terms and has the form
\begin{align*}
 J = & \sum_{iIk} c_k a_k(r_{iI}) \\
& + \sum_{ijm} c_m b_m(r_{ij}) \\ 
& + \sum_{ijIklm} c_{k l m} \Big\{ a_k(r_{i I}) a_l(r_{j I}) + a_k(r_{j I}) a_l(r_{i I}) \Big\} b_m(r_{ij})
\end{align*}
where the $a_k$ are one-body basis terms, the $b_m$ are two-body basis terms, the $c_k$, $c_m$, $c_{klm}$ are coefficients, and the $i,j$ label electrons while capital $I$ labels ions.
The sums in the Jastrow factor can be understood as taking into account electron-ion, 
electron-electron, and electron-electron-ion correlation, respectively~\cite{mi}.
Further details about the correlation functions can be found in Ref.\cite{ba}. 
 As explained above, we employ two types of trial functions. The first one has the HF nodes
so that
\begin{equation*}
\Psi_A=\Psi_{HF}
\end{equation*}
while the second state  
\begin{equation*}
\Psi_A= \Psi_{2conf}
\end{equation*}
Our one-particle orbitals were generated using a numerical HF code.  
The variational parameters to be optimized in variational Monte Carlo are the Jastrow coefficients, $c_k, c_m,$ and $c_{klm}$, 
a parameter in the basis terms $a_k$ and $b_m$, 
and the determinantal weight, $d_1$. These were optimized by algorithms which minimize the 
combination of energy and its variance as described elsewhere Ref.\cite{ba}.
All QMC calculations were done using the software package QWalk~\cite{wa}.

\subsection{Dependence of $E_{HF}$ and $E_{corr}$ on $Z$}
The correlation energy is defined as customary 
\begin{equation*}
{E}_{c}(N,Z) = {E}(N,Z)-E_{HF}(N,Z)
\end{equation*}
where $E(N,Z)$ is
 the total non-relativistic energy of an atom with nuclear charge $Z$  and 
$N$ electrons, while $E_{HF}(N,Z)$ is its Hartree-Fock counterpart.  
$E(N,Z) / Z^2$ can be expanded in a Laurent series as in~\cite{ch} 
\begin{align*}
E(N,Z) =& B_0(N) Z^2 + B_1(N) Z + B_2(N)\\
& + B_3(N) Z^{-1} + B_4(N) Z^{-2} + \ldots
\end{align*}
We can use 
M\o ller-Plesset perturbation theory to analyze the first few terms. In particular, one finds
that $E_{HF} = B_0(N) Z^2 + B_1(N) Z$.  The first term is a sum of the energies of occupied orbitals,
\begin{equation*}
B_0(N) Z^2 = \sum_{a} \epsilon_a
\end{equation*}
where each orbital energy $\epsilon_a$ is proportional to $Z^2$.   The next term is equal 
to first order correction to this energy with the perturbing potential being
 the difference between the exact potential and the effective Hartree-Fock potential, 
${\cal V}_{pert}=V_{Coulomb} - V_{HF}$,
\begin{equation*}
B_1(N) Z = \bra{\Psi_{HF}} {\cal V}_{pert} \ket{\Psi_{HF}}
\end{equation*}
which will be linear in $Z$.  For the four electron system, $E_{c}$ will 
increase linearly with $Z$~\cite{ch}~\cite{da}.  Chakravorty et. al. explain 
that this is due to the near-degeneracy effect since
the single configuration Hartree-Fock wave function is not the correct zeroth order wave function
in the non-interacting limit $Z^{-1}=0$. This deficiency means $\bra{\Psi_{HF}} {\cal V}_{pert} \ket{\Psi_{HF}}$ is not equal to $B_1$.  
As such, expansions of the HF energy and the actual total energy in Z$^{-1}$  for Be and Be-like atoms 
differ in the linear coefficient. This results
in a leading term proportional to $Z$ in the expression for $E_{corr}$~\cite{ch}.

\section{FNDMC results and discussion}
\begin{figure}
\includegraphics[width=\textwidth]{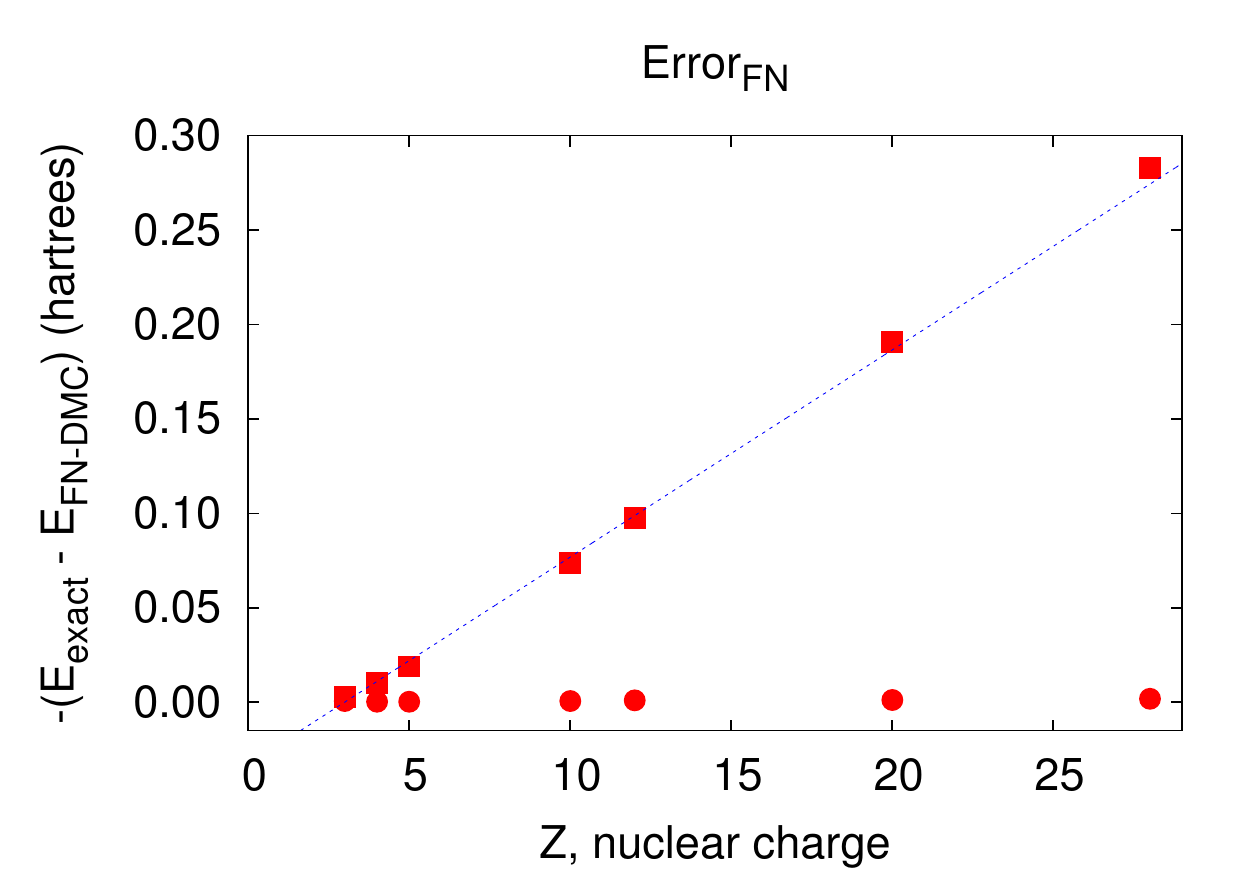}
\caption{A comparison of the FN-DMC error for different wave functions calculated 
using values in Table~\ref{table:CorrvsDensity}.  The squares correspond to the HF nodes while the circles correspond
to the 2-configuration nodes.  The linear fit to the error from the HF nodal structure has a slope of  0.0111(1).  The error bars are much smaller than the plot symbols.}
\label{fig:error}
\end{figure}

We carried out the fixed-node calculations for trial functions with $\Psi_{HF}$ and $\Psi_{2conf}$ nodes for
$Z=3, 4, 5, 10, 12, 20, 28$.  The results are listed in Table~\ref{table:CorrvsDensity} and plotted in Fig. \ref{fig:error}. The linear increase
with growing $Z$ is very clear and is quite striking considering that it covers the range from Li$^{-}$ through Be to
very highly ionized cases. Interestingly,
this linear increase matches the linear increase in the correlation energy based on perturbation analysis above.
In fact, the observed 
slope of -0.111(1) is very close the analytically derived result of -0.0117 
for the linear correction of the total energy in the MP2 theory \cite{ch}.

\begin{center}
\begin{table}
\centering     	    				
\caption{FN-DMC ground state energies for $\Psi_{HF}$ and $\Psi_{2conf}$  wave functions compared to the estimated exact
energies estimated from experiments~\cite{ch} for Z=4 through 28 and extrapolation~\cite{fi} to infinite basis set for Z=3 }
\begin{tabular}{l r r r} 		
\hline\hline
Species  		& -${\cal E}_{0}$	& -E$_{\Psi_{HF}}$	& -E$_{\Psi_{2conf}}$ \\[0.5ex]  
\hline 
Li$^{1-}$		&7.500758		&7.49812(6)		&7.5008(1)	\\
Be			&14.66736		&14.65715(4)		&14.66707(7)	\\
B$^{1+}$		&24.34893		&24.3300(2)		&24.3486(1)	\\
Ne$^{6+}$	&110.29092		&110.2167(2)		&110.2902(1)	\\
Mg$^{8+}$	&162.17108		&162.073(2)		&162.170(1)	\\
Ca$^{16+}$	&469.69474		&469.503(7)		&469.6935(1)	\\
Ni$^{24+}$	&937.21951		&936.936(3)		&937.217(9)	\\[1ex] 
\hline
\end{tabular}
\label{table:CorrvsDensity}
\end{table}
\end{center}

Increasing $Z$ from 3 to 28 has the effect of localizing the single particle orbitals radially with 
corresponding increase in the density closer to nucleus.  This can be 
measured by the expectation value of the radius for the orbitals $\langle r \rangle_{1s}$ and 
$\langle r \rangle_{2s}$ and each change by an order of magnitude (shown in Table~\ref{table:expectr}).  This clearly reveals that with increasing density the fixed-node error grows since the HF nodes 
are static and they do not depend on the density (we remind the HF node equation $(r_1-r_2)(r_3-r_4)=0$.  

\begin{table}
\caption{Expectation values of radius, r, for one particle numerical Hartree-Fock orbitals given in bohr}
\begin{center}
\begin{tabular}{|c|c|c|}
\hline
Z&		 $\langle r \rangle_{1s}$& 	$\langle r \rangle_{2s}$\\[0.5ex]
\hline
3&		0.572876& 				3.87342\\
4&		0.414896&				2.64902\\
5&		0.325176&				1.7977\\
10&		0.156101&				0.710569\\
12&		0.122388&				0.571132\\
20&		0.0765909&				0.32221\\
28&		0.0515183&				0.207193\\[1ex]
\hline
\end{tabular}
\end{center}
\label{table:expectr}
\end{table}

This is confirmed by the small fixed-node error of the $Z=3$ case, less than 4\% of $E_{corr}$, which is much
smaller than the error for Be which is larger than 10\% of $E_{corr}$. This is in line with the fact that
 the Li$^{-1}$  $2s$ electrons are highly diffuse and Jastrow-type  correlations  play much more important role
overall.
 
\begin{figure}
	\includegraphics[trim=4.75in 1in 6.75in 3.25in, clip, width=\textwidth]{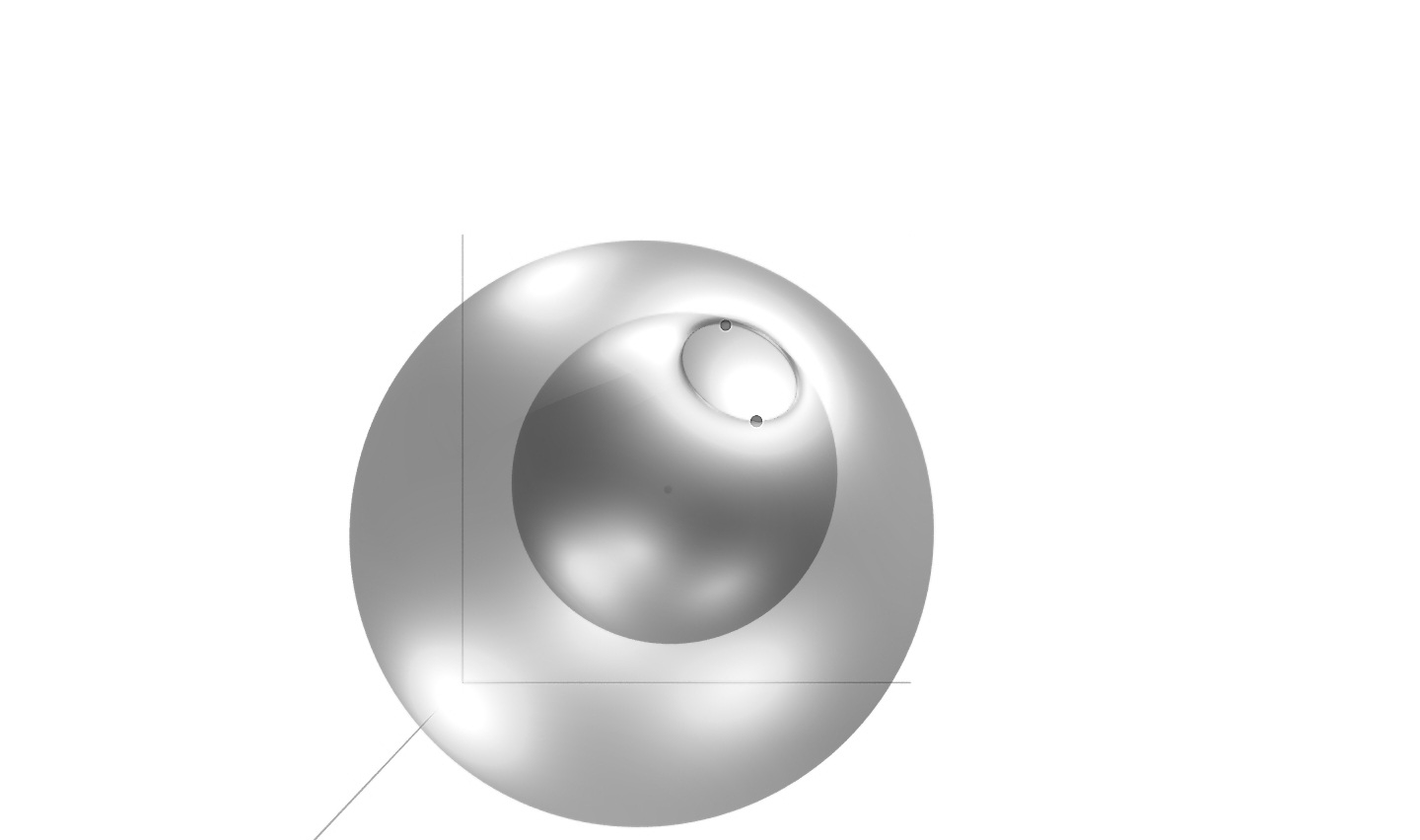}
	\caption{3D subspace of the 2-configuration nodal surface in real space.  
	The two dots at the opening represent 
the spin-up and -down electrons fixed at slightly different radial distances. The tiny dark spot 
in the middle is the nucleus. The node is found by scanning the space with the remaining two electrons located on the top
of each other and plotting the wave function's zero isosurface. 3 lighting sources are used to make the curvature of the surface visible.  The semi-transparency enables to see ``inside'' and show that the pair
of the scanning electrons can sample both inside and outside regions by passing through the opening 
(i.e., without crossing a node). This is not the case
for the HF wave function which has the nodal surface always as two concentric ideal spheres (one corresponding to spin-up
the other to spin-down subspaces). } 
	\label{fig:beciwf}
\end{figure}

One way to look at the nodes is to scan the wave function with a ``probe'' electron or pair of electrons
for a given position for the rest of the particles. One gets a  view of the subspace of the 11-dimensional node which
corresponds to our system. We therefore fixed one of the spin-up electrons  at $\lbrace r,\theta, \phi \rbrace
= 
\lbrace 1.0, \frac{7 \pi}{18},\frac{i\pi}{8} \rbrace$ and similarly, one of the spin-down electrons at
 $\lbrace 0.985, \frac{\pi}{4},\frac{\pi}{8} \rbrace$.  We then scanned the space by the remaining pair of electrons 
and we draw the zero isosurface of the $\Psi_{2conf}$, see Fig.
~\ref{fig:beciwf}. The scan shows clearly the effect of two configurations which enable the pair of the 
electrons to ``visit'' both inside and outside region of the two almost concentric spheres which are fused on one side
creating thus a nodal ``opening.'' This can be contrasted with the HF node which is given exactly by 
two ideal, concentric spheres. During the QMC walker evolution process  
the pair of particles thus cannot get from outside to inside--it will be forever locked 
 in one of the (four) equivalent domains. 

These results clearly indicate that the impact of the density on the fixed-node errors is very significant. In particular,
comparison of $Z=28$ with $Z=3$ is perhaps the most revealing. For $Z=28$ the electron density is highly localized 
around the nucleus, and the HF orbitals are very close to noninteracting ones since the electron-electron interaction
is much smaller than the electron-ion contribution.  However, the nodal error is significant both in absolute terms
(283 mHa) and in relative terms (28.6 \% of $E_{corr}$). Interestingly, this system is considered ``easy'' in the sense
that the one-particle contributions dominate overall. On the other hand, Li$^-$ exhibits nontrivial 
 many-body correlations related to the weak anionic bonding which is usually more demanding to describe by
traditional approaches based on basis sets and excitations. 
In fact, the FNDMC result is surprisingly good
even with the HF nodes, which, as explained above, have 
incorrect topology. This seems counterintuitive but can be understood
once we realize that significantly
lower density of Li$^-$ make the nodal errors less important. 

The second interesting observation is that the two-configuration wave function is very accurate and captures 
nodes correctly in the range of systems, from the weakly bonded anion to the very highly ionized cations. Although it
is easy to understand that the near-degeneracy effect is crucial it is interesting to see that higher
excitations appear to have only marginal effect for all studied systems.

\section{Conclusions}
We studied the dependence of the fixed-node error on electron density.  For this purpose we have chosen the system of four electrons in a Coulomb potential with varying nuclear charge $Z$. This system has somewhat unique properties:  the HF wave function's fixed-node bias is rather large, the HF node is invariant to varying $Z$, and for the Be atom it was known that very accurate nodes were not too difficult to construct.  Expanding on these insights, our calculations have convincingly shown:

a) the fixed-node bias increases with increasing density and appears to 
be predominantly linear with $Z$ and for the case of four-electron Be-like
Coulomb  system is very closely related to the first-order correction 
in the MP2 theory;  

b) the 2-configuration wave function produces highly accurate nodes across the values of $Z$ we tested,
i.e. it appears to be very consistent and basically universal for this system.  

This has important implications for the further analysis of the fixed-node errors for larger and more 
complicated systems since the results suggest that even small fixed-node errors in the region with 
high electronic density could have much stronger influence on the total correlation energy 
recovered than large fixed-node errors in low electronic density regions.

\section{Acknowledgements}
We would like to thank the Center for HIgh Performance Simulations (CHIPS) at North Carolina State University for computational resources, and also both Michal Bajdich and Jindrich Kolorenc for their insightful suggestions and comments.  I would like to thank the Physics Department and Graduate School of North Carolina State University for continuing support.
We are supported by DMR-0804549 and OCI-0904794 grants and by DOD/ARO and DOE/LANL
DOE-DE-AC52-06NA25396 grants. We
acknowledge also allocations at ORNL through INCITE and CNMS
initiatives as well as allocations at NSF, NCSA, and TACC centers.
% \appendix

% \section{}
% \label{}


\begin{thebibliography}{00}


% notes:

% \bibitem{label2}
% If there is a note, it should come last:
% \bibitem{label3} \note
% \end{subbibitems}

\bibitem{mi} W.~M.~C.~Foulkes,
L.~Mitas,
R.~J.~Needs,
G.~Rajagopal,
Rev. Mod. Phys. {\bf 73} (2001) 33.

\bibitem{br} D.~Bressanini,
D.~.M.~Ceperly,
P.~J.~Reynolds in:
W.~A.~Lester (Ed.),
Recent Advances in Quantum Monte Carlo Methods, World Scientific, Singapore, 2002, Pt. II, p. 1.

\bibitem{um} C.~J.~Umrigar,
M.~P.~Nightingale,
K.~J.~Runge,
J. Chem. Phys. {\bf 99} (1993) 2865.

\bibitem{um2} C.~J.~Umrigar,
K.~G.~Wilson, 
J.~W.~Wilkins, 
Phys. Rev. Lett. {\bf 60} (1998) 1719.

\bibitem{ba} M.~Bajdich,
 L.~Mitas,
 Acta Physica Slovaca {\bf 59} (2009) 81.
 
 \bibitem{wa} L.~K.~Wagner,
M.~Bajdich,
L.~Mitas,
J. Comp. Phys. {\bf 228} (2009) 3390.

\bibitem{ch} S.~J.~Chakravorty,
Steven~R.~Gwaltney,
E.~R.~Davidson,
F.~A.~Parpia,
C.~F.~Fischer,
Phys. Rev. A. {\bf 47} (1993) 3649.

\bibitem{da} E.~R.~Davidson,
S.~A.~Hagstrom,
S.~J.~Chakravorty,
V.~M.~Umar,
C.~F.~Fischer,
Phys. Rev. A. {\bf 44} (1991) 7071.

\bibitem{fi} C.~F.~Fischer,
J. Phys. B:  At. Mol. Opt. Phys. {\bf 26} (1993) 855.

\end{thebibliography}
\end{document}